# The Problems of Personal Income Tax on Revenue Generation in GOMBE State

Abubakar Bala[1], Esther Yusuf Enoch[2] & Salisu Yakubu[3]

[1]*department Of Business Administration Gombe State University, Nigeria*
[2 & 3]*department Of Banking And Finance Federal Polytechnic Mubi, Adamawa State, Nigeria*
*Corresponding Author: Abubakar Bala*

***Abstract:*** *This study examined the problems of personal income tax on revenue generation in Gombe state. The methodology used in data collection is survey, which utilized both primary and secondary types of data. Purposive sampling technique was adopted in selecting a sample of 150 respondents from both employees of state board of internal revenue service and tax payers in the state. The chi square statistics test was used in testing the hypotheses. The study found that tax avoidance/evasion and complete absences of information technology are serious problems affecting revenue generation in the state. It recommends that government should device strict measures in dealing and punishing individuals engage in tax avoidance and evasion. It should also employ the use of information technology as it is the only way problems experience in personal income tax collection can be reduced drastically.*

------------------------------------------------------------------------------------------------------------------

------------------------------------------------------------------------------------------------------------------

## I. INTRODUCTION

Most of the world's Countries economy either developed or developing is based on one form of taxation or the other. This is because tax is a very effective tool in generating revenue by governments as well as strengthens the economy. Worlu & Emeka (2012) opine that tax revenue utilization is a basis for supporting developmental activities in less developed economies. That is why different countries in the world today have different fiscal policies that enable them to explore various types of taxation and impose them on their citizens for the purpose of enhancing revenue generation so that to consolidate their economy. Therefore government of Nigeria, as one of those countries, through legislative powers has impose on its citizens forms of tax at whatever rate it deems appropriate. Similarly Gombe State government has its own legislative powers that allowed it to impose on its citizenry any form of tax and at whatever rate it deems appropriate so as to improved revenue generation ( NOUN, 2012).

As state earlier, the primary purpose of taxation is mainly to generate revenue for settling government expenditure, provision of social amenities and the welfare of the populace. Taxation is also used as an instrument of economic regulation for the purpose of discouraging or encouraging certain form of socioeconomic behavior. In addition it can be used to achieve specific economic objectives of nations. It is also a device to improve gross domestic product, induce economic development and influence favourable balance of payments with other countries (Magashula, 2010).

Perhaps the Nigerian economy relies solely on oil sector. This followed the discovering of oil in the country in the early 1970s. The situation had made both the Federal and State governments to neglect other sources of revenue generation to depend on oil to about 80-90% if not 99%. At the end of every month both state and federal will gathered at Abuja for monthly allocation derived from petroleum resources. However with the prevailing economic crunch orchestrated by dwindling fortunes from the oil and gas sector, governments at all levels are beginnings to adjust their expenditure profile, thereby looking for other sources of income to run their states. This is because the State's monthly allocation from the federation account has dropped and money had since stopped trickling in as before. Even the Sovereign Wealth Fund (SWF), which should have rescued the situation, was emptied long ago because the savings was shared between the states and Federal Government when the impact of free fall in oil price was beginning to be apparent at both level of government.

In order ameliorate the situation many states have started to unfold plans aimed at improving their Internal Generated Revenue (IGR). While some are planning to exploit mineral resources available within their domain, others are looking forward to attract direct investment to agriculture and tourism sectors. In Gombe State, the government has set up a committee to improve its Internal Generated Revenue (IGR) (Guardian News, 2015). Although this effort by Gombe State government continues to improve the state of revenue in the State, however, there are some problems confronting the process in the State. For instance, it is difficult to maximize tax revenue collection due to various forms of tax evasions. Eschborn (2010) observed that tax evasion is an issue that is conceivable as longstanding as taxation alike. Tax evasion occurs when people or organizations deliberately fail to abide by their tax responsibility (Simser, 2008). Irrespective of its values, tax evasion denies every government the tax revenue due to the system, which results in a gap between the potential and actual tax collection (Adebisi & Gbegi, 2013). Perhaps personal income tax as one of the components of Internal Generated Revenue (IGR) in Gombe State as well as Nigeria in general has remained the most unsatisfactory, problematic and disappointing of all taxes over a long period of time in today's tax system. This emanate as a result of poor data management, lack of tax information and complete absence of information technology which led to the tax avoidance, tax evasion and non-remittance of some tax generated by tax officials in to the government coffers in the State. Tax evasion and tax avoidance result in loss of tax revenues and impair the chances of realizing the distributional or equity goal of taxation, and, if they become widespread, as they have in recent times, then more tax payers may lose faith in the tax administration system and may be tempted to join the ranks of tax evaders (Piketty, 2014).

Although several studies revealed and identified so many problems affecting personal income tax on revenue generation in several part of the world and Nigeria in particular. However studies related to problems affecting personal income tax on revenue generation in Gombe state are rare or scarce. This creates a gap that this study (The problems of personal income tax on revenue generation in Gombe state) intends to fill.

**Objective of the Study**
The main objective of this study is to assess the problems of personal income tax on revenue generation in Gombe State. The specific objectives were to;
 i. examine the effect of tax avoidance and evasion on revenue generation in Gombe State.
 ii. examine the effect of absence of information technology on revenue generation in Gombe State.

## II. METHODOLOGY

The methodology of conducting the study was survey design. It involved the use of both primary and secondary types of data. Primary data allowed the researchers to gather the opinion of individuals (tax collection staff and tax payer) associate with problems of personal income tax on revenue generation in Gombe state. Purposive sampling was used in selection of 150 respondents in the state. The data collected was analyzed and tested with chi square test.

## III. LITERATURE REVIEW
**Personal Income Tax**

Personal income tax is a tax levied on all incomes of individuals employed by a business or organization either public or private. Self employed individuals are expected to file annual returns annually and pay the taxes due. A tax payer is required to file returns for the proceeding year within 90 days of the end of the year. Personal income tax could be; Direct Assessment Tax, Self Assessed Tax and Pay as You Earn (PAYE) Tax. Direct assessment is used to assess tax for self employed individuals from their personal income. However the self assessed tax, a new tax payer can assess him/herself, pay the calculated tax at designated banks and get e-TCC (Tax Clearance Certificate) without visiting any tax office. Pay as You Earn (PAYE) taxes implies all employers in Nigeria are responsible for deducting earning (Pay As You Earn) from their employees' pay. Taxes deducted from earning are required to be remitted to the appropriate tax office by the 10th day of the month following the deduction. This depicts that monthly payment of Pay As You Earn (PAYE) tax liabilities are to be made on or before the 10th day of the month following the applicable month (e.g. January tax to be remitted by 10th of February).

**Tax Avoidance/Evasion and Revenue Generation**

Tax avoidance is the use of legal methods to modify an individual's financial situation to lower the amount of income tax owed. This is generally accomplished by claiming the permissible deduction and credits. As a form of tax non-compliance, tax avoidance is an entity that intends to subvert a Gombe State tax system thereby reducing its internal revenue generation. In other hand tax evasion is the amount of unreported income, which is the difference between the amount of income that should be reported to the tax authorities and the actual amount reported. Tax evasion is as a result of loopholes in the tax laws, poverty and lack of adequate public enlightenment which deprived government adequate revenue generation in the State. This implies that tax avoidance and evasion is a great threat to the health or wealth of revenue generation in the state. Tax avoidance is a threat to revenue generation in Gombe state because it allowed individuals to escape payment of tax or to pay some little portion.

**Absence of Information Technology on Revenue Generation**

Application of information technology has become increasingly necessary in Nigeria's tax system and revenue generation. This is because the use of information technology makes for fast, easy and accurate computation, storage and presentation/retrieval of data/records of tax payers. Therefore, in order to improve on the efficiency of tax administration and revenue generation in Gombe state, information and technology must be considered rather than manually compiled database ( use of files/folders for data storage) of tax payers). Absence of Information Technology on Revenue Generation can lead to tax avoidance and evasion. That is why Gombe State tax system remained paralyzed as a result of poor data management of personal income tax payer as a result of complete absence of information technology to keep proper record of personal income tax payers. Absence of Information Technology on Revenue Generation led to the tax avoidance, evasion and non-remittance of some tax generated by tax officials in to the government coffers in the State thereby leading to great setback or drawback in revenue generation in the state..

**Revenue Generation**

Revenue Generation is a complete amount of money that is generated during a specific time or period. Government revenue includes all amounts of money received from sources outside the Governments entity. The sources include revenue generated from individuals which include personal income tax. It is a normal tradition that Governments at all levels usually have an agency or department responsible for collecting Government revenue from companies and individuals. In Gombe State for example, Internal Revenue Service Board is responsible for collecting taxes on behalf of Government at the State level.

**International Issues on the Study**

International monetary fund survey shows that tax policy can help spur Economic growth of nations through; Strong tax collection system that help to raise revenue for governments (IMF, 2012). Growth-friendly objectives must be balanced with improved distribution of income. Well-designed tax policy improves prospects for economic growth. However, growth objectives must be balanced with those to promote an improved distribution of income, a regional tax conference heard in Brasilia, Brazil.

Most Latin America economies continue to perform well, despite many economic uncertainties elsewhere in the world. Yet across Latin America, policy makers continue to work to find ways to boost economic growth."Strengthening fiscal frameworks requires building stronger and more reliable revenue bases," said IMF Deputy Managing Director Naoyuki Shinohara, who delivered opening remarks at the conference in Brazil. "This is necessary to help countries weather global shocks—and limit income volatility that comes with increased commodity dependence— but also sustain growth over the medium term. A stronger revenue base would provide a more stable source of income to finance much needed public investment in the region. Better designed tax structures could encourage growth while promoting equity. "Focusing on the structure of the tax system, Gilbert Terrier, deputy head of the IMF's Fiscal Affairs Department, emphasized that a key challenge in the region will be to increase the revenue productivity and structure of the personal income tax so that it can contribute to growth- enhancing public expenditure while helping attain distributional goals. Continued efforts toward improving the effectiveness of tax administration in the region would help bolster tax revenues, the IMF said. Countries in the region with relatively more effective tax administrations also have the best tax revenue performance.

Building tax administration capacity is needed to help spur development in Africa. A new survey shows that action is being taken, but more work is needed. Tax revenues account for over a third of GDP in OECD countries. But they account for far less in developing countries, particularly in sub-Saharan Africa, where they correspond to less than a fifth of GDP. More tax revenue would not only help the governments of these countries function and pay for goods and services, but would open the way for other market and state reforms that would promote economic, social and environmental development. Raising tax burdens might seem like an odd proposition to policymakers in crisis stricken OECD countries as they bid to raise revenues while keeping tax burdens as light as they can for the sake of growth. But when taxes account for 10 to 15% of GDP, a well-designed increase in tax is exactly what many developing countries need: just as an excessively heavy tax burden might crush activity, an excessively low one can starve an economy of the oxygen it needs to advance. But how can more tax revenue be raised in poorer economies? One way is to generate more growth, but as many of the countries concerned lack the resources to administer tax, this approach may not be enough, that is, not unless efforts are made to improve the effectiveness of tax administration systems at the same time. That means strengthening the capacity and resources needed for better taxpayers' services and enforcement, reviewing tax structures, and investing in skills and management systems needed to produce corruption-free tax systems. This is easier said than done, of course, but improving the effectiveness and transparency of tax administration systems is nonetheless widely accepted as a key step to achieving the UN Millennium Development Goals. And because it means mobilizing more domestic taxation, it can also help in smoothing efforts to open up world trade by further reducing the reliance on border taxes. In other words, channeling funds, including development aid, towards better tax administration is money well spent. Indeed, as OECD data shows, tax/GDP ratios in sub-Saharan countries where tax administration reforms are being implemented now exceed 16.8% of GDP, which was the average for fragile and lower income countries. But to make a bigger difference, more information is needed about how tax administrations actually work and where the problems lie. To fill this gap, the International Tax Dialogue, a global initiative based at the OECD and involving the EU, the IMF and the World Bank, among others, has undertaken a survey of 15 sub- Saharan African revenue bodies–Benin, Botswana, Burundi, Ethiopia, Ghana, Kenya, Malawi, Mauritius, Rwanda, Senegal, Sierra Leone, South Africa, Tanzania, Uganda and Zambia (OECD, 2011). The aim is to build a clear picture as to the various approaches and practices used across the continent, to identify problems and to provide policymakers with a better view of the kinds of measures that might be taken to address them. Similar work has already been carried out for the 50 middle and higher income countries of the OECD's Forum on Tax Administration. The good news is that all of the countries surveyed by the International Tax Dialogue are currently engaged in some pretty significant tax administration reforms, often with donor support. Nevertheless, our pilot survey has revealed some instructive trends and patterns. Take cost, which is one of the main challenges facing tax administrators in developing countries. The cost of collection varies from 1% to 4% of the total collected in the region. Salary and related expenditures account for the largest portion–some 60- 80% of the budget. In most of the surveyed countries, investment in information technology accounts for less than 2% of total administrative expenditure. It should therefore come as no surprise to learn that most of the revenue bodies surveyed reported being dissatisfied with their existing IT systems. Efforts are being made to address this: all revenue bodies surveyed, with the exception of Burundi, indicated that they have a separate and substantive in-house IT function, and some are developing or plan to implement integrated tax administration systems for self-help services such as online registration, filing and payment. Further investment in administrative systems would undoubtedly help improve treatment of taxpayers too. As the survey shows, all revenue bodies, except South Africa, assign identification numbers ostensibly across all tax types, including customs Ekpo & Akpan ,1994),. All personal and corporate income tax systems are based on self-assessment principles. VAT is a feature across all countries surveyed, with a few countries using two thresholds, one for the sale of goods and another for services. In fact, indirect taxes contribute the highest proportion of revenue in seven of the countries surveyed, with direct taxes in six countries and international trade taxes in two countries. Non-tax revenues such as income from state-owned enterprises, fees and other payments for government services account for a very small proportion–about 1 to 2%–of total revenue collection. Compare this to developing countries in Latin America, where these can reach 10% or more of government revenues. Regarding enforcement, all countries believe they have adequate powers to enforce the payment of tax, with various interest and penalty regimes for similar offences applying across tax types, specifically for income taxes on the one hand and for VAT on the other (GCC, 2015). However, it is unclear how effective these measures are in practice: this is an area for further exploration. Institutional arrangements are another issue which can have an impact on the effectiveness of tax administration. The revenue bodies in most of the countries surveyed follow a relatively unified, semi-autonomous model, meaning that they have considerable freedom to interpret tax laws, allocate resources, design internal structures and implement appropriate human resource management strategies. At the same time, they are responsible for tax, customs and non-tax revenue operations. Three of the countries surveyed are now integrating the collection of social security contributions with tax operations, a trend that is emerging in OECD countries too. As for organisational arrangements, most are hybrid in nature. In line with current practice in tax administration, a number of revenue bodies have set up a headquarters function to provide operational policy guidance to fi eld delivery. Moreover, all revenue bodies (except Botswana) have set up a "large taxpayers' office" to administer all tax affairs of major enterprises and some individuals. Apart from Botswana and Mauritius, the revenue bodies have also created special taxation regimes for small and micro-enterprises, and six countries have set up dedicated units to manage them. Meanwhile, all revenue bodies surveyed produce 3-5 year business/corporate plans using established planning frameworks, with clear mission statements, visions and objectives, as well as the actions to reach them, as do OECD countries. Most of the revenue bodies are funded through parliamentary appropriations, meaning that they develop budget proposals and bid for funding just like any other government department or agency. Some countries provide their revenue body with a performance bonus, such as a percentage of the collections, which is a practice rarely found elsewhere. As this pilot survey suggests, real efforts are evidently underway to build effective tax administrations in several African countries, which is good news for long-term economic development. However, the devil is in the detail, and a more comprehensive study is to be carried out in collaboration with the African Tax Administration Forum (ATAF) and other international institutions to paint a clearer picture of the character and particular needs of the various administrations across the region. This information will also feed into the G20-led push on tax and development. That means covering more countries, and collecting and further refining the data, while drawing comparisons with countries outside the region. It is a major undertaking, but if countries in sub-Saharan Africa can use the information to help them improve their tax policies and their development paths at the same time, then the task will be worth it (Alan, 2015).

**The Nigerian Context**
The Personal Income Tax (Amendement) Act 2011 amended Personal Income Tax Cap. P8 Laws of the Federation of Nigeria, 2004. The amendment became effective June 14, 2011 from the day it was signed into Law by President Goodluck Jonathan. Sadly, the gazette was only made public January 2012 with the possibility that taxes computed or paid on the extant Act made require some readjustment or refunds for periods prior to January 2012 since they amendment became effective June 14, 2012. The likelihood of tax credits cannot be ruled out as refund is not easy to obtain in the Nigerian tax environment even though a law permits same. Therefore this study presents the key highlights that would have an impact on revenue generation in Gombe state and Nigeria by large from the effective date. However relief and allowance were previously exempted.

**Relief and Allowance:** Section 33(1): Consolidated Relief Allowance of the higher N200, 000 plus 20% of gross income and 1% of gross income. This clearly state that the balance shall be taxable after all relief and exemptions have been deducted accordance with the Income Table in the Sixth Schedule (Onyere, 2008). Further the following deductions are tax exempt; National Housing Fund Contribution, National Health Insurance Scheme, Life Assurance Premium, National Pension Scheme and Gratuities. However the New tax table (Effective June 14. 2011) revealed that; First N300,000 taxed at 7%, Next N300,000 taxed at 11%, Next N500,000 taxed at 15%, Next N500,000 taxed at 19%, Next N1,600,000 taxed at 21% and Above N3,200,000 taxed at 24%.

Meanwhile Section 33(2) defines gross emoluments as wages, salaries, allowances (including benefits in kind, gratuities, superannuation and any other incomes derived solely from employment (Orewa, 2004). Section 36(6) empowers the Minister to assess income tax on a tax payers under a presumptive tax regime based on order stated in gazette for that purpose. This suggest a continuation of the Best of Judgment Assessment regime where the tax payer makes its practically impossible for the revenue service to do a proper ascertainment of income and assess tax accordingly. Section 52 prescribe a penalty of N50,000 for Individuals and N500,000 for Corporate Entities where the taxable person fails or refuses to keep books of account which in the opinion of the relevant tax authority are adequate for the purpose of the tax assessments.

**Compulsory Tax Deductions:** Section 74(1) makes it compulsory for tax payers to deduct tax on rent(10%), interest(10%), director fees(10%) and Dividend(10%) and remit same to the tax authorities within 30days the amount was deducted or the time the duty to deduct arose. Failure to do this is punished with a penalty of 10% of tax not deducted or remitted in addition to the amount of tax or remitted plus interest at the prevailing monetary policy rate of the Central Bank of Nigeria (Ola, 2004).

**Interest for Late Tax Payment:** Section 77 Interest penalties on late payment of taxes will now be on an annual basis from due date till it is settled.

**Pay As You Earn:** Section 81(2) require every employer to file a return with the relevant tax authority of all emolument paid to its employees, not later than 31 January of every year in respect of all employees in its employment in the preceding year. Section 81(3) prescribes a penalty on conviction of N500, 000 in the case of a corporate entity and N50, 000 in the case of an individual where Section 81 (2) is Breached the government of Gombe state and Nigeria at large as power to disdain. Power to disdain is clearly spelt in Section 104(1) which requires the relevant tax authority to obtain a court order to disdain a defaulting tax payer (Stephen, 2011)..

**Tax Clearance Certificate**: Section 85 require Ministries, Department or an agency of the Government or a Commercial Bank to verify the genuineness of Tax Clearance Certificate presented to it by a tax payer from the relevant issuing tax authorities. Further, they should require Tax Clearance when there is change of ownership of vehicles, application for plot of land or any other transaction as may be determined from time to time (Taiwo, 2002).

**Concluding Thought:** The new law expects tax payers to be more compliant more than ever before. In addition, there is likely to a reduction in income taxes to the relevant tax authorities due to the increase in the tax relief and exemption particularly for low income earners that constitute the bulk of tax payers. It however see tax authorities making up this shortfall with pursuit of penalties for contravention of relevant sections of the law (Mathew, 2013).

**Empirical Study**

Several studies have been conducted on problems of Personal Income Tax. The studies carried out were hinged on, the problems of Personal income tax collection and administration on revenue generation in Gombe State and Nigeria in general. However this study relies solely on the problems of personal income tax on revenue generation in Gombe state. Therefore some of these empirical studies were however presented below to aid in discussing the findings of this study.

Maryam (2011) conducted a study on Tax collection and Revenue Generation in Gombe State. The study aimed to identify the problems associated with associated collection in the State. Simple random sampling technique was used and 51 questionnaires were distributed to the staff of Gombe state internal revenue service and some selected taxpayers in the state. The findings of the research shows that, tax avoidance, evasion, bribe and corruption are the major problems affecting revenue generation in the state.

Nwabor (2014) assessed the problems and prospects of personal income tax administration in Nigeria. The study was set up to look at the problems in assessing, collecting and administering Personal income tax in Nigeria. The study adopted simple random sampling in selecting 73 respondents. The outcome of the research shows that there are problems associated with personal income tax collection and administration in the state. Finally the researcher recommend that; enlightenment campaign and staff training, simplicity of tax assessment, quality internal control system, record keeping and computerization to be improved by the tax regulators so as to enhance its role in revenue generation for the state.

Al'amin (2011) carried out a research on problems of Nigeria personal income tax collection and administration. The study found that, lack of appropriate tax laws, unskilled personnel, poor data management and mismanagement of tax collected are responsible for the problems associated with personal income tax on revenue generation in Nigeria, and, it can only be minimized when appropriate measures are put in place.

**Theoretical Framework**

Economists have put forward many theories/models or principles of taxation at different times to guide the state as to how justice or equity in taxation can be achieved. Among the theories are; Benefit Theory, The cost of Service Theory, Ability to Pay Theory and Proportionate Principle.

**Benefit Theory:** According to this theory, the state should levy taxes on individuals according to the benefit conferred on them. The more benefits a person derives from the activities of the state, the more he should pay to the government. This principle has been subjected to severe criticism on the following grounds;

Firstly, if the state maintains a certain connection between the benefits conferred and the benefits derived. The tax payer can be able to pay taxes and vice versa. This theory is against the basic principle of the tax. That tax is mandatory and compulsory contribution made by citizenry to the public authorities to meet the expenses of the government and the provisions of general benefit. There is no direct quid pro quo in the case of a tax.

Secondly, most of the expenditure incurred by the state is for the general benefit of its citizens. It is not possible to estimate the benefit enjoyed by a particular individual every year. Therefore aligning payment of tax with benefit derived is not justifiable.

Thirdly, if to apply this principle in practice, then the poor will have to pay the heaviest taxes, because they benefit more from the services of the state. If we get more from the poor by way of taxes, it is against the principle of justice, equity and fairness.

**The cost of Service Theory**

Some economists were of the opinion that if the state charges actual cost of the service rend from the people, it will satisfy the idea of equity or justice in taxation. The cost of service principle can no doubt be applied to some extent in those cases where the services are rendered out of prices and are a bit easy to determine, e.g. postal, railway services, supply of electricity etc. But most of the expenditure incurred by the state cannot be fixed for each individual because it cannot be exactly determined. For instance, how can it measure the cost of service of the police, armed forces, judiciary, etc., to different individuals? Dalton has also rejected this theory on the ground that there s no quid pro qua in a tax. In other hand the theory is seem to be the opposite of benefit theory that advocate for citizen to pay taxes on basis of the benefit derived from the authority collecting the tax.

**Ability to Pay Theory**

The most popular and commonly accepted principle of equity or justice in taxation is that citizens of a country should pay taxes to the government in accordance with their ability to pay. It appears very reasonable and just taxes should be levied on the basis of the taxable capacity of an individual. For instance, if the taxable capacity of a person A is greater than the person B, the former should be asked to pay more taxes than the latter. It seems that if the taxes are levied on this principle as stated above, then justice can be achieved. But our difficulties do not end here. The fact is that when we put this theory in practice, our difficulties actually begin. The trouble arises with the definition of ability to pay. The economists are not unanimous as to what should be the exact measure of a person's ability or faculty to pay. The main viewpoints advanced in this connection are as follows;

Ownership of Property: Some economists are of the opinion that ownership of the property is a very good basis of measuring one's ability to pay. This idea is out rightly rejected on the ground that if a person earns a large income but does not spend on buying any property, he will then escape taxation. On the other hand, another person earning less income buys property; he will be subjected to taxation. Is this not absurd and unjustifiable that a person, earning large income is exempted from taxes and another person with small income is taxed?

Tax on the Basis of Expenditure: It is also asserted by some economists that the ability or faculty to pay tax should be judged by the expenditure which a person incurred. The greater the expenditure, the higher should be the tax and vice versa. The viewpoint is unsound and unfair in every respect. A person having a large family to support has to spend more than a person having a small family. As the test of one's ability to pay, the former person who is already burdened with many dependents will have to' pay more taxes than the latter who has a small family. So this is unjustifiable.

Income as the Basics: Most of the economists are of the opinion that income should be the basis of measuring a man's ability to pay. It appears very just and fair that if the income of a person is greater than that of another, the former should be asked to pay more towards the support of the government than the latter. That is why in the modern tax system of the countries of the world, income has been accepted as the best test for measuring the ability to pay off a person.

**Proportionate Principle**

In order to satisfy the idea of justice in taxation, J. S. Mill and some other classical economists have suggested the principle of proportionate in taxation. These economists were of the opinion that if taxes are levied in proportion to the incomes of the individuals, it will extract equal sacrifice. The modern economists, however, differ with this view. They assert that when income increases, the marginal utility of income decreases. The equality of sacrifice can only be achieved if the persons with high incomes are taxed at higher rates and those with low income at lower rates. They favor progressive system of taxation, in all modern tax systems.

Looking from all angles or corners regarding the theories explained above proportionate theories is best to be adopted by Gombe state government. This is because the theory is on the opinion that taxes are to be levied in proportion to the incomes of the individuals, so as to extract equal sacrifice.

**Data Analysis and Presentation**

As earlier stated in the methodology the method of data analysis adopted in the study nonparametric statistics (chi square). The models used enable the study to examine the problems of personal income tax on revenue generation in Gombe state

**Test of Hypothesis**

The two hypotheses that were formulated earlier in the beginning of the study was tested using chi-square technique.

Table 1 $Ho_1$: Tax avoidance and evasion has no significant effect on revenue generation in Gombe State.

| O | E | O - E | $(O – E)^2$ | $(O – E)^2/E$ |
|---|---|---|---|---|
| 18 | 24.67 | -6.67 | 44.49 | 1.80 |
| 25 | 16.67 | 8.33 | 69.39 | 4.16 |
| 5 | 5.67 | -0.67 | 0.44 | 0.08 |
| 2 | 3 | -1 | 1 | 0.33 |
| 0 | 0 | 0 | 0 | 0 |
| 30 | 24.67 | 5.33 | 28.41 | 1.15 |
| 16 | 16.67 | -0.67 | 0.44 | 0.03 |
| 3 | 5.67 | -2.67 | 7.13 | 1.26 |
| 1 | 3 | -2 | 4 | 1.33 |
| 0 | 0 | 0 | 0 | 0 |
| 26 | 24.67 | 1.33 | 1.77 | 0.07 |
| 9 | 16.67 | -7.67 | 58.83 | 3.53 |
| 9 | 5.67 | 3.33 | 11.09 | 1.96 |
| 6 | 3 | 3 | 9 | 3 |
| 0 | 0 | 0 | 0 | 0 |
|   |   |   |   | 18.7 |

**Source:** Researcher's Computation, 2018.

The chi square table value can be calculated through DF= (C-1) (R-1), therefore DF= (5-1) (3-1) DF= (4) (2) DF= 8, 5% under 8 =15.51

The chi square calculated value is 18.7 ($X^2 = 18.7$)

**Decision:** Since the calculated chi-square value (18.7) is greater than the chi-square table value (15.51) (18.7>15.51), the null hypothesis is rejected while the alternative hypothesis is accepted that is tax avoidance and evasion has significant effect on revenue generation in Gombe state..

**Table 2 Ho$_2$:** Absence of information technology affect revenue generation in Gombe State

| O | E | O - E | $(O – E)^2$ | $(O – E)^2/E$ |
|---|---|---|---|---|
| 16 | 13.67 | 2.33 | 5.43 | 0.40 |
| 20 | 2.4 | -4 | 16 | 0.67 |
| 3 | 8.33 | -5.33 | 28.41 | 3.41 |
| 11 | 4 | 7 | 49 | 12.25 |
| 0 | 0 | 0 | 0 | 0 |
| 9 | 13.67 | -4.67 | 21.81 | 0.67 |
| 28 | 2.4 | 4 | 16 | 0.67 |
| 12 | 8.33 | 3.67 | 13.47 | 1.62 |
| 1 | 4 | -3 | 9 | 2.25 |
| 0 | 0 | 0 | 0 | 0 |
| 16 | 13.67 | 2.33 | 5.43 | 0.40 |
| 24 | 2.4 | 0 | 0 | 0 |
| 10 | 8.33 | 1.67 | 2.79 | 0.33 |
| 0 | 4 | 0 | 0 | 0 |
| 0 | 0 | 0 | 0 | 0 |
| | | | | **22.01** |

**Source:** Researcher's Computation, 2018.

The table value can be calculated through DF= (C-1) (R-1), therefore DF= (5-1) (3-1) DF= (4) (2) DF= 8, 5% under 8 =15.51

The calculated value is $X^2 = 22.01$

**Decision:** Since the calculated chi-square value (22.01) is greater than the chi-square table value (15.51) (22.01>15.51), the null hypothesis is rejected while the alternative hypothesis is accepted that is Absence of information technology has significant effect on revenue generation in Gombe State.

## IV. DISCUSSION AND FINDINGS OF THE STUDY

The study was set to assess the problems of Personal income tax on revenue generation. In order to assess the problem, two specific objectives were set to examine the problems of; tax avoidance and evasion on revenue generation in Gombe state and absence of information technology on revenue generation in Gombe state. It was found that;

Tax avoidance and evasion has significant effect on revenue generation in Gombe state. The findings are in collaboration with Maryam (2011) findings that tax avoidance and evasion are serious problem that affect revenue generation. This implies that tax avoidance and evasion was found to be a serious threat affecting revenue generation of personal income tax in the state. A large number of tax payers avoid and evade paying their personal income tax by either lowering the amount of income owed or decline to report their actual income to the officials responsible for tax collection in the state.

Absence of information technology has significant effect on revenue generation in Gombe state. This was buttressed by Nwabor (2014) and Al,amin (2011) that record keeping and computerization to be improved by the tax regulators so as to enhance its role in revenue generation for the state. This illustrates that absence of information technology affect revenue generation of personal income tax negatively as revealed by the large number of respondents. Tax officials find it difficult to store, present/retrieve and compute accurate information of tax payers obtained as a result of absence of information technology in the state tax system.

## V. CONCLUSION AND RECOMMENDATIONS

**Conclusion**

The study found out that tax avoidance, evasion and absence of information technology in Gombe State is problems that affect personal income tax on revenue generation. That is why the tax system in the state remained paralyzed as a result of tax avoidance and evasion as well as complete absence of information technology which is a serious threat to the State. Poor revenue generation in the state has led it to rely solely on Federal government monthly allocations which keep on fluctuation thereby leading to uncertainty. Therefore the problems of personal income tax on revenue generation in the state require radical handling to ensure that a large chunk of taxable populace does not escape it. Thus, an effective and efficient tax system will help ensured the problems of personal income tax on revenue generation in Gombe State are minimized.

## RECOMMENDATIONS

i. The state government should employ rigid and strict measures in dealing and punishment of those individuals who try to avoid evade tax. The individuals might emanate from tax officials or individual who suppose to pay the tax or both.
ii. Tax authorities in the state should adopt the use of modern information technology in keeping the tax payer's record so as to have accurate information of tax payers and avoid delay in tracing and retrieving tax taxpayer's record so that large number of personal income tax payers would not escape from paying their taxes
iii. Tax avoidance and evasion: It appears that taxes should be levied on the basis or ability of an individual to pay as the only way tax avoidance and evasion of personal income tax can be minimized in the state. Also, tax avoidance and evasion can be reduced when the revenue generated from personal income tax payers is used for the welfare of the general populace.
iv. Absence of information technology: There is need for the tax authorities to employ the use of information technology as the only way problems experience in personal income tax collection can be reduced drastically. Training and retraining of personal income tax officials regarding the use of information technology in tax collection could also help to rescue the situation in time
v. Government should engage in enlightenment campaign, staff training and good tax policy would help tax authorities in dealing with tax offenders and encourage tax payers in the state.